\documentclass[aps,pra,reprint,showpacs]{revtex4-1}
\usepackage{amsmath}
\usepackage{SIunits}
\usepackage{graphicx}

\begin{document}
%
%
%
\renewcommand{\a}[1]{\ensuremath{\hat{a}_{#1}}}
\newcommand{\ac}[0]{\ensuremath{\hat{a}_{\mathrm{c}}}}
\newcommand{\acprime}[0]{\ensuremath{\hat{a}_{\mathrm{c}}'}}
\renewcommand{\ap}[0]{\ensuremath{\hat{a}_{\mathrm{p}}}}
\newcommand{\aR}[0]{\ensuremath{\hat{a}_{\mathrm{R}}}}
\newcommand{\aT}[0]{\ensuremath{\hat{a}_{\mathrm{T}}}}
\newcommand{\adag}[1]{\ensuremath{\hat{a}_{#1}^{\dagger}}}
\newcommand{\adagc}[0]{\ensuremath{\hat{a}^{\dagger}_{\mathrm{c}}}}
\newcommand{\adagcprime}[0]{\ensuremath{\hat{a}_{\mathrm{c}}'^{\dagger}}}
\newcommand{\adagp}[0]{\ensuremath{\hat{a}^{\dagger}_{\mathrm{p}}}}
\newcommand{\adagR}[0]{\ensuremath{\hat{a}_{\mathrm{R}}^{\dagger}}}
\newcommand{\adagT}[0]{\ensuremath{\hat{a}_{\mathrm{T}}^{\dagger}}}
\newcommand{\alphac}[0]{\ensuremath{\alpha_{\mathrm{c}}}}
\newcommand{\alphaLO}[0]{\ensuremath{\alpha_{\mathrm{LO}}}}
\newcommand{\betain}[0]{\ensuremath{\beta_{\mathrm{in}}}}
\newcommand{\bra}[1]{\ensuremath{\left<#1\right|}}
\renewcommand{\c}[0]{\ensuremath{\hat{c}}}
\newcommand{\cdag}[0]{\ensuremath{\hat{c}^{\dagger}}}
\renewcommand{\d}[0]{\ensuremath{\hat{d}}}
\renewcommand{\ddag}[0]{\ensuremath{\hat{d}^{\dagger}}}
\newcommand{\deff}[0]{\ensuremath{\hat{d}_{\mathrm{eff}}}}
\newcommand{\Deltaa}[0]{\ensuremath{\Delta_{\mathrm{q}}}}
\newcommand{\Deltaaeff}[0]{\ensuremath{\Delta_{\mathrm{q}}^{\mathrm{eff}}}}
\newcommand{\Deltaca}[0]{\ensuremath{\Delta_{\mathrm{cq}}}}
\newcommand{\Deltac}[0]{\ensuremath{\Delta_{\mathrm{c}}}}
\newcommand{\dens}[0]{\ensuremath{\hat{\rho}}}
\newcommand{\etaeff}[0]{\ensuremath{\eta_{\mathrm{eff}}}}
\newcommand{\Fmin}[0]{\ensuremath{F_{\mathrm{min}}}}
\newcommand{\Fopt}[0]{\ensuremath{F_{\mathrm{opt}}}}
\newcommand{\Gammameas}[0]{\ensuremath{\Gamma_{\mathrm{m}}}}
\newcommand{\gammap}[0]{\ensuremath{\gamma_{\mathrm{p}}}}
\newcommand{\gammapar}[0]{\ensuremath{\gamma_{\parallel}}}
\newcommand{\gammapareff}[0]{\ensuremath{\gamma_{\parallel}}^{\mathrm{eff}}}
\newcommand{\gammaperp}[0]{\ensuremath{\gamma_{\perp}}}
\newcommand{\gammaperpeff}[0]{\ensuremath{\gamma_{\perp}}^{\mathrm{eff}}}
\newcommand{\gp}[0]{\ensuremath{g_{\mathrm{p}}}}
\renewcommand{\H}[0]{\ensuremath{\hat{H}}}
\newcommand{\Heff}[0]{\ensuremath{\hat{H}_{\mathrm{eff}}}}
\renewcommand{\Im}[0]{\ensuremath{\mathrm{Im}}}
\newcommand{\kappap}[0]{\ensuremath{\kappa_{\mathrm{p}}}}
\newcommand{\ket}[1]{\ensuremath{\left|#1\right>}}
\renewcommand{\L}[0]{\ensuremath{\mathcal{L}}}
\newcommand{\mean}[1]{\ensuremath{\langle#1\rangle}}
\newcommand{\omegaa}[0]{\ensuremath{\omega_{\mathrm{q}}}}
\newcommand{\omegac}[0]{\ensuremath{\omega_{\mathrm{c}}}}
\newcommand{\omegaL}[0]{\ensuremath{\omega_{\mathrm{L}}}}
\newcommand{\Omegalockin}[0]{\ensuremath{\Delta_l}}
\newcommand{\pauli}[0]{\ensuremath{\hat{\sigma}}}
\newcommand{\pexc}[0]{\ensuremath{p_{\mathrm{exc}}}}
\newcommand{\psuccess}[0]{\ensuremath{p_{\mathrm{suc}}}}
\renewcommand{\Re}[0]{\ensuremath{\mathrm{Re}}}
\renewcommand{\S}[0]{\ensuremath{\hat{S}}}
\newcommand{\tauc}[0]{\ensuremath{\tau_{\mathrm{c}}}}
\newcommand{\taujitter}[0]{\ensuremath{\tau_{\mathrm{jit}}}}
\newcommand{\taulockin}[0]{\ensuremath{\tau_l}}
\newcommand{\Tr}[0]{\ensuremath{\mathrm{Tr}}}
\newcommand{\vac}[0]{\ensuremath{\hat{v}}}
\newcommand{\Var}[0]{\ensuremath{\mathrm{\Var}}}
\renewcommand{\vec}[1]{\ensuremath{\mathbf{#1}}}
\newcommand{\X}[0]{\ensuremath{\hat{X}}}
\newcommand{\zetalockin}[0]{\ensuremath{\zeta_{\mathrm{lockin}}}}
\newcommand{\zetalockinthr}[0]{\ensuremath{\zeta_{\mathrm{lockin}}^{\mathrm{thr}}}}
\newcommand{\zetameanI}[0]{\ensuremath{\zeta_{\mean{I}}}}
\newcommand{\zetameanIthr}[0]{\ensuremath{\zeta_{\mean{I}}^{\mathrm{thr}}}}

\title{Measurement-induced two-qubit entanglement in a bad cavity: \\
  Fundamental and practical considerations}

\author{Brian Julsgaard}
\email{Electronic mail: brianj@phys.au.dk}
\author{Klaus M{\o}lmer}
\affiliation{Lundbeck Foundation Theoretical Center for Quantum System
  Research, Department of Physics and Astronomy, Aarhus University, Ny
  Munkegade 120, DK-8000 Aarhus C, Denmark.}


\date{\today}

\begin{abstract}
  An entanglement-generating protocol is described for two qubits
  coupled to a cavity field in the bad-cavity limit. By measuring the
  amplitude of a field transmitted through the cavity, an entangled
  spin-singlet state can be established probabilistically. Both
  fundamental limitations and practical measurement schemes are
  discussed, and the influence of dissipative processes and
  inhomogeneities in the qubits are analyzed. The measurement-based
  protocol provides criteria for selecting states with an infidelity
  scaling linearly with the qubit-decoherence rate.
\end{abstract}

\pacs{03.67.Bg, 42.50.Pq, 42.50.Lc, 03.65.Ta}

\maketitle

\section{Introduction}
\label{sec:introduction}
Entanglement is one of the key features of quantum mechanics, and
during the past decades it has been demonstrated experimentally in
many different physical systems. By coherent control of interacting
quantum systems, entangled states can be engineered directly
\cite{Hagley.PhysRevLett.79.1(1997), Sackett.Nature.404.256(2000),
  Schmidt-Kaler.Nature.422.408(2003), Liebfried.Nature.422.412(2003),
  Ansmann.Nature.461.504(2009),
  DiCarlo.Nature.460.240(2009)}. Alternatively, entanglement can be
established as a consequence of the outcome of a measurement process
--- either as a continuous (possibly quantum-non-demolition (QND))
measurement \cite{Julsgaard.Nature.413.400(2001)} or as a consequence
of a single quantum jump \cite{Chou.Nature.438.828(2005),
  Moehring.Nature.449.68(2007)}. Some of the above examples employ
cavity quantum electrodynamics (QED)
\cite{Kimble.PhysicaScripta.T76.127(1998)} for mediating the
interaction between the quantum systems, which allows for the direct
engineering of entangled states in the strong-coupling regime
\cite{Hagley.PhysRevLett.79.1(1997), Ansmann.Nature.461.504(2009),
  DiCarlo.Nature.460.240(2009)}. The present paper considers a
different case --- the bad-cavity limit of QED, in which the damping
rate of the cavity field is fast compared to the coupling rate between
qubits and the cavity field. Hence, any information of qubit coherence
being encoded into the cavity field will immediately be lost from the
cavity and the above-mentioned direct-engineering schemes are
inapplicable. However, turning to a measurement-based protocol, the
detection of a field transmitted through the cavity allows for
re-establishing a firm knowledge of the qubit state and hence for the
creation of entangled states through measurement back action. The
measurement is of the continuous type, which is theoretically
well-described by stochastic master-equation methods
\cite{Jacobs.ContemporaryPhys.47.279(2006)}. In contrast to the work
of Ref.~\cite{Hutchison.CanJPhys.87.225(2009)} using similar
theoretical methods, our calculations are not restricted to the
dispersive and linear regime of the coupling between the cavity field
and the qubits but allow instead for a more generalized set of
parameters (even a resonant coupling) in search for the optimal choice
for entanglement generation. We consider feasible experimental
approaches and discuss the physical limitations imposed both
fundamentally by the measurement process and practically by
decoherence mechanisms.

This paper is arranged as follows: The basic idea for the protocol is
outlined in Sec.~\ref{sec:The_baisc_idea}, while the theoretical
modeling is elaborated on in Sec.~\ref{sec:stochastic_ME}. Various
practical measurement schemes are analyzed in
Sec.~\ref{sec:ent-generation-no-decay}, while qubit decoherence and
inhomogeneities are added in
Sec.~\ref{sec:Decoherence_processes}. After a general discussion in
Sec.~\ref{sec:discussion}, we summarize the conclusions of the paper
in Sec.~\ref{sec:Conclusion}. Some mathematical details are deferred
to the appendix.
\section{Entanglement-generating protocols: The basic idea} 
\label{sec:The_baisc_idea}
\begin{figure}
  \centering
  \includegraphics[width=\linewidth]{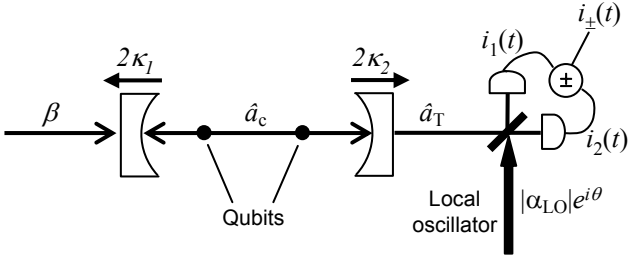}
  \caption{The physical system under consideration. Two qubits are
    coupled to a cavity field, $\ac$, which is driven externally by a
    constant coherent field, $\beta$ (in the frame rotating at the
    driving frequency, $\omegaL$). The field-decay rates through the
    cavity mirrors are denoted by $\kappa_1$ and $\kappa_2$. The
    output field, $\aT$, is subjected to a balanced homodyne
    measurement using a local oscillator, $\alphaLO =
    |\alphaLO|e^{i\theta}$. The differential and summed photo-currents
    are denoted by $i_-(t)$ and $i_+(t)$, respectively.}
\label{fig:CavitySetup}
\end{figure}

The physical setup under consideration (see
Fig.~\ref{fig:CavitySetup}) consists of two qubits placed in a cavity
subjected to an external driving field and to a continuous measurement
by employing a phase-sensitive detection of the field leaking from the
cavity. Let the ground and excited states of either qubit be denoted
by $\ket{\mathrm{g}}$ and $\ket{\mathrm{e}}$, respectively, and
consider the two-qubit basis set $\{\ket{\mathrm{ee}},
\ket{\mathrm{gg}}, \ket{+}, \ket{-}\}$, where $\ket{\pm} =
\frac{1}{\sqrt{2}}(\ket{\mathrm{eg}} \pm \ket{\mathrm{ge}})$. Our aim
is to generate the spin-singlet state, $\ket{-}$, by a probabilistic
detection scheme with a high fidelity and a high success
probability. This state does not couple to the cavity field when the
coupling parameter $g$ is equal for the two qubits. The other states,
$\ket{\mathrm{ee}}$, $\ket{\mathrm{gg}}$, and $\ket{+}$ span the
spin-triplet space and the cavity field may induce rotations within
this subset of Hilbert space. Furthermore, the coupling between the
qubits and the cavity field gives rise to a correlated decay
mechanism, which induces transitions (with rate $2\gammap$) within the
triplet space: $\ket{\mathrm{ee}} \rightarrow \ket{+}$ and $\ket{+}
\rightarrow \ket{\mathrm{gg}}$, while the singlet state, $\ket{-}$, is
unaffected.

The idea is now to prepare a separable initial state, $\ket{\psi}$,
and subsequently to collapse (probabilistically) $\ket{\psi}$ into
$\ket{-}$ by the measurement process. The separable initial states of
opposite spins, $\ket{\mathrm{eg}} = \frac{1}{\sqrt{2}}(\ket{+} +
\ket{-})$ or $\frac{1}{2} (\ket{\mathrm{g}} + \ket{\mathrm{e}})
(\ket{\mathrm{g}} - \ket{\mathrm{e}}) = \frac{1}{2}\ket{\mathrm{gg}} -
\frac{1}{2}\ket{\mathrm{ee}} + \frac{1}{\sqrt{2}}\ket{-}$, both have a
50\% overlap with the desired singlet state, and our task is to
identify a measurement scheme, which is able to distinguish between
the singlet and triplet components of $\ket{\psi}$ and thus to
facilitate the state collapse.

No real or virtual transitions can take place within the one
dimensional singlet subspace of the qubits and it experiences no
interaction with the cavity field. The triplet subspace, however,
consists of three states, and their interaction with the cavity field
can be tailored to affect the transmitted radiation field in two
different ways: the cavity field and the damping of the spin system
via the cavity mode can drive the collective spin to a steady state
mean spin polarization which causes a phase shift of the transmitted
radiation, or the dynamical driving of the triplet spin components can
induce a frequency modulation of the transmitted field
auto-correlation function. The steady state change in the transmitted
field is visible in the homodyne photo-current, while the frequency
modulation can be observed with a lock-in detector.
\section{The stochastic-master-equation approach}
\label{sec:stochastic_ME}
Our knowledge of the physical system depicted in
Fig.~\ref{fig:CavitySetup} is accounted for by the density matrix,
$\dens$, which evolves according to the stochastic master equation
\cite{Jacobs.ContemporaryPhys.47.279(2006),
  Wiseman.PhysRevA.47.642(1993)}:
\begin{equation}
\label{eq:MasterEq_original}
  \frac{\partial\dens}{\partial t} = \frac{1}{i\hbar}[\H,\dens] 
    +\sum_m\mathcal{D}[\c_m]\dens + \sqrt{\eta}\mathcal{H}[\d]\dens\xi(t),
\end{equation}
where the Hamiltonian, $\H$, describes the interaction between the
cavity field, the qubits, and the coherent driving field, $\betain =
\beta e^{-i\omegaL t}$. Decay processes are modeled by the
super-operator, $\mathcal{D}[\c_m]\dens = -\frac{1}{2}\cdag_m\c_m\dens
-\frac{1}{2}\dens\cdag_m\c_m + \c_m\dens\cdag_m$, while our knowledge
from the continuous monitoring of the system is incorporated by the
measurement super-operator, $\mathcal{H}[\d]\dens = \d\dens +
\dens\ddag - \mean{\d+\ddag}\dens$. The detector quantum efficiency is
denoted by $\eta$, and the real-valued function, $\xi(t)$, models the
randomness of the detection process with ensemble characteristics,
$\mean{\xi(t)}_{\mathrm{E}} = 0$ and
$\mean{\xi(t)\xi(t')}_{\mathrm{E}} = \delta(t-t')$. In the frame
rotating at the driving frequency, $\omegaL$, the Hamiltonian reads:
\begin{align}
  \H &= \hbar\Deltac\adagc\ac 
    +i\hbar\sqrt{2\kappa_1}(\beta\adagc - \beta^*\ac) 
  \notag \\ & \quad + \frac{\hbar\Deltaa}{2}\S_z
  +\hbar g(\S_+\ac + \S_-\adagc),    
  \label{eq:Hamiltonian_general}
\end{align}
where $\Deltac = \omegac-\omegaL$ and $\Deltaa = \omegaa - \omegaL$
denote the detuning of the driving frequency, $\omegaL$, from the
cavity and qubit resonance frequencies, $\omegac$ and $\omegaa$,
respectively, and $g$ is the coupling strength between light and
qubits. The cavity-field creation and annihilation operators are
denoted by, $\adagc$ and $\ac$, respectively, while $\S_k =
\sum_{j=1}^2\pauli_k^{(j)}$ for $k = +,-,z$ are sums of Pauli
operators, $\pauli_k^{(j)}$, for the two qubits. The coherent driving
amplitude, $\beta$, is normalized such that $|\beta|^2$ is the
incident number of photons per second onto the input mirror, the
field-decay rate of which is $\kappa_1$. Similarly, with $\kappa_2$
being the field-decay rate of the exit mirror, the leakage of the
cavity field is modeled by the decay operator, $\c_1 =
\sqrt{2\kappa}\ac$, in the decay part of
Eq.~(\ref{eq:MasterEq_original}), where $\kappa = \kappa_1 +
\kappa_2$. Population decay of qubit 1 and 2 with rate, $\gammapar$,
can be modeled by $\c_2 = \sqrt{\gammapar}\pauli_-^{(1)}$, $\c_3 =
\sqrt{\gammapar}\pauli_-^{(2)}$, respectively, whereas collision-like
phase decay of each dipole moment is modeled by $\c_4 =
\frac{1}{\sqrt{2\tau}}\pauli_z^{(1)}$ and $\c_5 =
\frac{1}{\sqrt{2\tau}}\pauli_z^{(2)}$, where $\tau$ is the mean
waiting time between the phase-disrupting events. The output field is
given by \cite{Collett.PhysRevA.30.1386(1984)} $\aT =
\sqrt{2\kappa_2}\ac -\vac$, where the vacuum field, $\vac$, reflected
from the exit mirror preserves the operator commutation relations but
gives no further contribution at zero temperature (methods for
treating finite-temperature environments are outlined in
Ref.~\cite{Julsgaard.PhysRevA.85.013844(2012)}). The balanced homodyne
detection setup mixes the output field and the local oscillator field,
$\alphaLO$, leading to the differential
\cite{Wiseman.PhysRevA.47.642(1993)} and summed photo-currents (in
units of electrons per second):
\begin{equation}
  \begin{split}
  i_-(t) &= \sqrt{\eta}|\alphaLO|\left[2\sqrt{2\eta\kappa_2}
    \mean{\X_{\theta}(t)} + \xi(t)\right], \\
  i_+(t) &= \eta|\alphaLO|^2,  
  \end{split}
\end{equation}
where the field-quadrature operator, $\X_{\theta} = \frac{1}{2}(\ac
e^{-i\theta} + \adagc e^{i\theta})$, depends on the relative phase,
$\theta$, of the local oscillator. The operator $\X_{\theta}$ is
connected to the formalism of Eq.~(\ref{eq:MasterEq_original}) when
$\d$ is defined by $\d = \sqrt{2\kappa_2}\ac e^{-i\theta}$, i.e.~$\d +
\ddag = 2\sqrt{2\kappa_2}\X_{\theta}$. By defining the normalized
differential photo-current, $I(t) \equiv i_-(t)/\sqrt{i_+(t)}$, we
obtain:
\begin{equation}
\label{eq:NormalizedCurrent_and_X}
  I(t) = 2\sqrt{2\kappa_2\eta}\mean{\X_{\theta}(t)} + \xi(t)
       = \sqrt{\eta}\mean{\d + \ddag} + \xi(t).
\end{equation}
With the notation, $\mean{A,B} \equiv \mean{AB}-\mean{A}\mean{B}$, the
correlation function, $R(t;\tau) =
\mean{I(t+\tau),I(t)}_{\mathrm{E}}$, of the normalized differential
photo-current is given by \cite{Wiseman.PhysRevA.47.642(1993)}:
\begin{equation}
  R(t;\tau) = 8\kappa_2\eta\mean{:\!\X(\theta,t+\tau),\X(\theta,t)\!:} 
    +\delta(\tau),
\end{equation}
where ``:'' means normal-ordering of the field operators.
\subsection{Adiabatic elimination of the cavity-field variables}
\label{sec:adiab-elim-cavity}
The above dynamical equations are very general and can be simplified
in our case of $\kappa \gg g$ by adiabatically eliminating the
cavity-field variables. Our elimination procedure varies only slightly
from previous works (see e.g.~\cite{Wang.PhysRevA.71.042309(2005)})
and hence only the main steps are given: The cavity-field operator is
written as $\ac \equiv \alphac + \acprime$, where $\alphac =
\frac{\sqrt{2\kappa_1}\beta}{\kappa+i\Deltac}$ corresponds to the mean
cavity-field in absence of qubits. Next, transform the master equation
to the frame rotating at the qubit resonance frequency, $\omegaa$, and
eliminate adiabatically $\acprime$. The resulting master equation is
then transformed back to the frame rotating at $\omegaL$ with the
effective qubit Hamiltonian given by:
\begin{equation}
\label{eq:Heff}
  \Heff = \frac{\hbar\Deltaa}{2}\S_z + \hbar g(\alphac\S_+ + \alphac^*\S_-)
  -\frac{\hbar\Deltaca g^2 \S_+\S_-}{\kappa^2 + \Deltaca^2},
\end{equation}
where $\Deltaca = \Deltac-\Deltaa$. The qubit-decay operators, $\c_m$
(with $m = 2,\ldots,5$), in the master
equation~(\ref{eq:MasterEq_original}) are maintained while the
cavity-leakage operator is replaced by the correlated qubit operator,
$\c_1 \rightarrow \sqrt{\gammap}\S_-$, where $\gammap =
\frac{2g^2\kappa}{\kappa^2 + \Deltaca^2}$. The measurement operator,
$\d$, is replaced by:
\begin{equation}
\label{eq:deff}
  \deff = \sqrt{2\kappa_2}\left(\alphac - \frac{ig\S_-}{\kappa+i\Deltaca}
    \right)e^{-i\theta},
\end{equation}
which in turn from Eq.~(\ref{eq:NormalizedCurrent_and_X}) leads to the
photo-current:
\begin{equation}
\label{eq:I(t)_ad_elim}
  \begin{split}
  I(t) =  - \sqrt{\gammap\etaeff}&[\mean{\S_x}\sin(\theta-\theta_{\kappa}) 
    +\mean{\S_y}\cos(\theta-\theta_{\kappa})] \\
   2\sqrt{2\kappa\etaeff}&[\Re\{\alphac\}\cos\theta 
    +\Im\{\alphac\}\sin\theta] + \xi(t),
  \end{split}
\end{equation}
where $\tan\theta_{\kappa} = -\frac{\Deltaca}{\kappa}$, $\S_{\pm} =
\frac{1}{2}(\S_x \pm i\S_y)$, and $\etaeff =
\frac{\eta\kappa_2}{\kappa}$ is the total detection efficiency
accounting also for the non-detected fraction,
$\frac{\kappa_1}{\kappa}$, of photons leaking through the left-hand
mirror in Fig.~\ref{fig:CavitySetup}. The photo-current correlation
function turns into:
\begin{equation}
\label{eq:R(t;tau)_ad_elim}
  \begin{split}
    R(t;\tau) = \gammap\etaeff[&\langle\S_+(t+\tau),\S_-(t)\rangle +  
     \langle\S_+(t),\S_-(t+\tau)\rangle \\
   -&\langle:\!\S_+(t),\S_+(t+\tau)\!:\rangle e^{2i(\theta-\theta_{\kappa})} \\
   -&\langle:\!\S_-(t+\tau),\S_-(t)\!:\rangle e^{-2i(\theta-\theta_{\kappa})}]
     + \delta(\tau),
  \end{split}
\end{equation}
where the normal-ordering is transferred from $\ac$, $\adagc$ to
$\S_-$, $\S_+$.

When inserted into the master equation~(\ref{eq:MasterEq_original}),
the two left-most terms in the Hamiltonian~(\ref{eq:Heff}) together
with the qubit-decay terms given by $\c_2,\ldots,\c_5$ corresponds
exactly to the semi-classical description of light-matter
interactions. In addition, the presence of the cavity introduces a
Stark-shift term (right-most term in Eq.~(\ref{eq:Heff})) and an
additional, correlated spontaneous decay process by the
$\c_1$-operator (the so-called Purcell effect
\cite{Purcell.PhysRev.69.681(1946)}). The elimination of the cavity
field is a good approximation whenever $\kappa \gg g,\chi,\gammapar,
\tau^{-1}$, where $\chi = 2g\alphac$ is the resonant Rabi frequency of
the qubits. For the remaining part of this manuscript we assume the
adiabatic elimination of the cavity field to be in effect when
referring to the master equation~(\ref{eq:MasterEq_original}).

This parameter regime is relevant for transmission-wave-guide
resonators \cite{Kubo.PhysRevLett.105.140502(2010),
  Schuster.PhysRevLett.105.140501(2010),
  Amsuss.PhysRevLett.107.060502(2011)}, in which the coupling of $N$
electronic spins has reached the strong-coupling regime, $\sqrt{N}g
\gg \kappa,\gammapar,\tau^{-1}$, while typical values for the coupling
parameter, $g/2\pi$, to a single spin could be extended to, say, 300
Hertz. As we shall learn in Sec.~\ref{sec:Decoherence_processes}, the
qubit-decoherence rate must be significantly smaller than $\gammap$,
which could be realized by coupling, e.g., single atomic ions
\cite{Olmschenk.PhysRevA.76.052314(2007)} to such wave guides.

\subsection{Optimal strategy for entanglement generation}
\label{sec:Optimal_strategy}
\begin{figure}
  \centering
  \includegraphics[width=\linewidth]{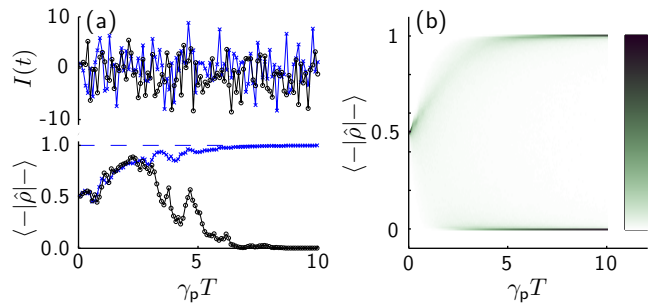}
  \caption{(Color online) (a) The photo-current (upper part) and the
    singlet-state overlap (lower part) as a function of time for two
    individual instances of the simulations. One (blue crosses)
    eventually collapses into $\ket{-}$ whereas the other (black
    circles) collapses into the triplet space. Panel (b) shows for
    each integration time, $T$, the distribution of the singlet-state
    overlap on a relative scale set by the shaded bar.}
  \label{fig:BayesianExamples}
\end{figure}
The stochastic master equation~(\ref{eq:MasterEq_original})
establishes the connection between, on one side, our knowledge of the
quantum state described mathematically by $\dens$, and on the other
side, the measurement record, $I(t)$. This connection can be
established in two different ways. (I) From a theoretical perspective,
Eq.~(\ref{eq:MasterEq_original}) presents a tool for simulating the
realizations of measurements. The stochastic function, $\xi(t)$, is
then generated by the computer software and gives rise to the
particular instance of $\dens(t)$ and eventually the photo-current
given by Eq.~(\ref{eq:I(t)_ad_elim}). (II) From an experimental
perspective, the stochastic master equation can be employed for
analyzing real experiments in which the photo-current, $I(t)$, has
been measured. The function $\xi(t)$ then represents the randomness of
the measurement process, and by continuously updating the coupled
equations~(\ref{eq:MasterEq_original}) and~(\ref{eq:I(t)_ad_elim}),
the density matrix, $\dens(t)$, will always correspond to the best
obtainable knowledge of the two-qubit quantum state. Even though the
present work employs method (I) for simulating the
entanglement-generating process, the full access to $\dens$ gives the
possibility to judge how well method (II) would work in experiment.

The evolution of $\dens$ and $I(t)$ during the measurement process is
exemplified in Fig.~\ref{fig:BayesianExamples}. For now we focus on
the qualitative features and we defer a discussion of the specific
physical parameters to our detailed presentation of results in
Sec.~\ref{sec:ent-generation-no-decay}. In panel (a) two instances of
the simulations have been shown; one which collapses into $\ket{-}$
(blue crosses), and one which does not (black circles). Despite the
fact that the two photo-current examples are both quite noisy, they do
contain enough information in order to increase the knowledge of the
singlet-state overlap, which eventually becomes zero or unity as shown
in the lower part of panel (a).  In panel (b) the
singlet-state-overlap distribution is shown versus time (based on
10,000 simulations). This overlap is initialized at 50\% but soon
attains a much broader distribution.  However, after few times
$\gammap^{-1}$ of measurement, the overlap-distribution bifurcates
into sharp peaks at zero and unity. Hence, for a sufficiently long
measurement time the continuous evolution of $\dens$ effectively
facilitates the desired wave-function collapse.

The optimal entanglement-generating protocol simply uses $\dens$ to
check to which degree the singlet state, $\ket{-}$, has been
realized. By requiring a minimum value, $\Fmin$, for the state
overlap, the acceptance criterion for a given quantum state then
becomes $\mean{-|\dens|-} \ge \Fmin$.
\subsection{Practical strategy for entanglement generation}
\label{sec:Practical-strategy}
The optimal strategy discussed above requires knowledge of all the
physical parameters, $\kappa_1$, $\kappa_2$, $\Deltac$, $\Deltaa$,
$g$, $\gammapar$, $\tau$, in addition to sufficient data processing
capability. This is indeed possible but might be impractical in
reality, and hence some more robust but less accurate procedures for
establishing whether the singlet state has been prepared are
desired. To this end we shall consider the two integrated,
dimensionless measurement signals:
\begin{align}
\label{eq:zetameanI}
  \zetameanI &= \frac{1}{\sqrt{T}}\int_0^T I(t) dt, \\
\label{eq:zetalockin}
  \zetalockin &= \frac{2}{T\taulockin}\int_0^T \left|\int_0^t 
    I(t')e^{-(i\Omegalockin+\frac{1}{\taulockin})(t-t')}dt'\right|^2 dt.
\end{align}
In experiment, the former of these corresponds to a simple integration
of the photo-current, while the latter corresponds to inserting the
photo-current signal into a lock-in amplifier with demodulation
frequency, $\Omegalockin$, and time constant, $\taulockin$, and
integrating for the measurement time, $T$, the squared modulus-output
value, $R^2 = X^2 + Y^2$ ($X$ and $Y$ are the measured in-phase and
in-quadrature amplitudes of the signal at frequency, $\Omegalockin$).
Now, our task is to devise conditions, e.g., $|\zetameanI| \le
\zetameanIthr$ or $\zetalockin \le \zetalockinthr$, to accept the
quantum state as being sufficiently close to $\ket{-}$. When
simulating the entire entanglement-generating process, the procedure
is repeated $N_{\mathrm{total}}$ times, and if $N_{\mathrm{accept}}$
of these simulation runs lead to acceptance of the quantum state, we
define the success probability as $\psuccess =
N_{\mathrm{accept}}/N_{\mathrm{total}}$. At the same time, the
fidelity $F = \frac{1}{N_{\mathrm{accept}}}\sum \bra{-}\dens\ket{-}$
measures the average occupation of the spin-singlet state for the
generated quantum states, where the sum runs over the accepted density
matrices, $\dens$, simulated by the master
equation~(\ref{eq:MasterEq_original}). In principle, it should be
possible to reach $F=1$ with $\psuccess = \frac{1}{2}$ (since the
initial state has a 50\% overlap with $\ket{-}$). However, in practice
a finite measurement time, qubit decoherence, and a non-optimal
extraction of information from $I(t)$ reduces the fidelity obtained at
a given $\psuccess$. An acceptance criterion is well-chosen if both
$F$ and $\psuccess$ attain high values.

\section{Entanglement generation in absence of qubit decay}
\label{sec:ent-generation-no-decay}
This section is devoted to the generation of the spin-singlet state in
absence of qubit-population and qubit-phase decay as modeled by the
decay operators $\c_2,\ldots,\c_5$, i.e.~we take $\gammapar = 0$,
$\tau = \infty$. This simplifies the introduction of all the detailed
concepts in the measurement scheme and defines the limits imposed
solely by the measurement setup and by the chosen acceptance
criteria. The influence of qubit decay is discussed in
Sec.~\ref{sec:Decoherence_processes}.

In the numerical simulations the stochastic part of
Eq.~(\ref{eq:MasterEq_original}) is integrated by the Milstein formula
\cite{Kloeden.StochasticDifferentialEquations}, and $\dens$ is evolved
using time steps, $dt$, being $10^{-2}$ times the characteristic decay
time or oscillation time of the physical variables. By repeating the
simulations 10,000 times, the statistical spread on fidelity estimates
is of the order of one percent.

Without loss of generality, the phase of the driving field, $\beta$,
can be chosen such that $\alphac$ and $\chi$ are real. We shall also
take $\Deltac = 0$, i.e.~$\beta$ must then be real. By taking
$\kappa_1 \approx 0$ and $\kappa_2 = \kappa - \kappa_1 \approx \kappa$
(i.e.~the entire cavity decay takes place through the right-hand
mirror in Fig.~\ref{fig:CavitySetup}) the effective quantum
efficiency, $\etaeff$, corresponds to that of the detector, $\etaeff
\approx \eta$. In experiment such a mirror asymmetry is not
necessarily realistic, but adding a homodyne detection setup to the
left-hand mirror output and combining the knowledge from all
measurements would re-establish $\etaeff \approx \eta$, and hence the
choice $\kappa_1 \approx 0$ just simplifies the simulations while
maintaining the experimental realism. The narrow qubit linewidth calls
for $\Deltaa = \Deltaca \ll \kappa$, and the correlated decay rate
becomes $\gammap \approx \frac{2g^2}{\kappa}$. In the remaining part
of this manuscript, all rates are measured relative to $\gammap$ and
time is measured in units of $\gammap^{-1}$. For the simulations we
take specifically $\kappa = 5000\gammap$ and $g = 50\gammap$, such
that $\frac{g}{\kappa} = \frac{1}{100}$ ensures the validity of the
adiabatic elimination.

\subsection{Measurement schemes}
\label{sec:measurement-schemes}

\subsubsection{DC-analysis of the photo-current}
\label{sec:dc-analysis}
\begin{figure}
  \centering
  \includegraphics[width=\linewidth]{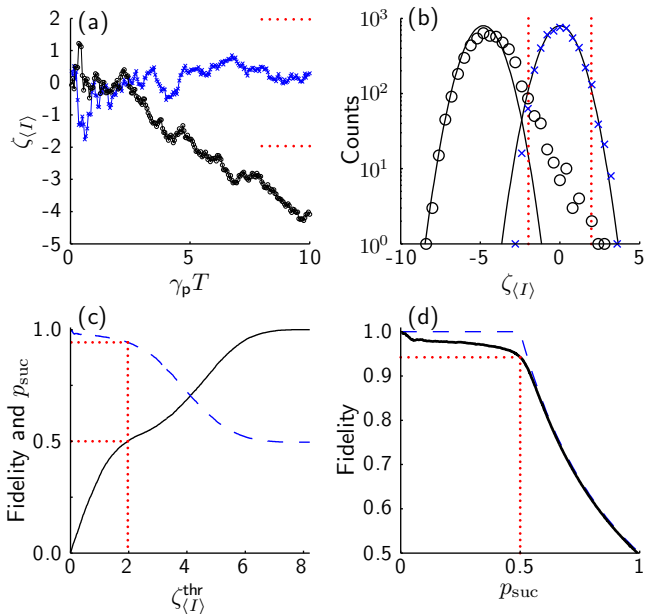}
  \caption{(Color online) All panels correspond to $\Deltaa =
    10\gammap$, $\chi = 1.65\Deltaa$, $\theta = -\pi/2$, and $\eta =
    1$. (a) Two instances of the integrated measurement signal,
    $\zetameanI$, as a function of the integration time, $T$. These
    examples are based on the photo-currents shown in
    Fig.~\ref{fig:BayesianExamples}(a), which led to the singlet state
    (blue crosses) or the triplet state (black circles). The
    red-dotted lines mark a chosen threshold condition, $|\zetameanI|
    \le \zetameanI^{\mathrm{thr}} = 1.96$, separating accepted and
    rejected states after the integration time, $\gammap T=10$. Panel
    (b) shows from 10,000 simulations the distribution of the
    measurement signal, $\zetameanI$, in case of $\mean{-|\dens|-} \ge
    0.8$ (crosses) and $\mean{-|\dens|-} \le 0.2$ (circles). The solid
    lines show Gaussian distributions with unit variance and with mean
    values given by the approximate estimate of
    Eq.~(\ref{eq:Ensemble_mean_zetameanI}). The acceptance window from
    panel (a) is marked with red-dotted lines. (c) The obtained
    fidelity, $F$ (dashed line), and success probability, $\psuccess$
    (solid line), as a function of the threshold value,
    $\zetameanI^{\mathrm{thr}}$, of the acceptance window. The
    red-dotted lines correspond to the choice of panels (a,b) leading
    to $\psuccess = 50\%$ and $F = 94\%$. (d) Solid line: The
    fidelity, $F$, as a function of the success probability,
    $\psuccess$. The red-dotted line marks $\psuccess =
    50\%$. Blue-dashed line: Theoretical limit for completely separate
    singlet- and triplet-space-measurement signals and perfect state
    overlap.}
  \label{fig:DCSimulationExample}
\end{figure}
\begin{figure}
  \centering
  \includegraphics[width=\linewidth]{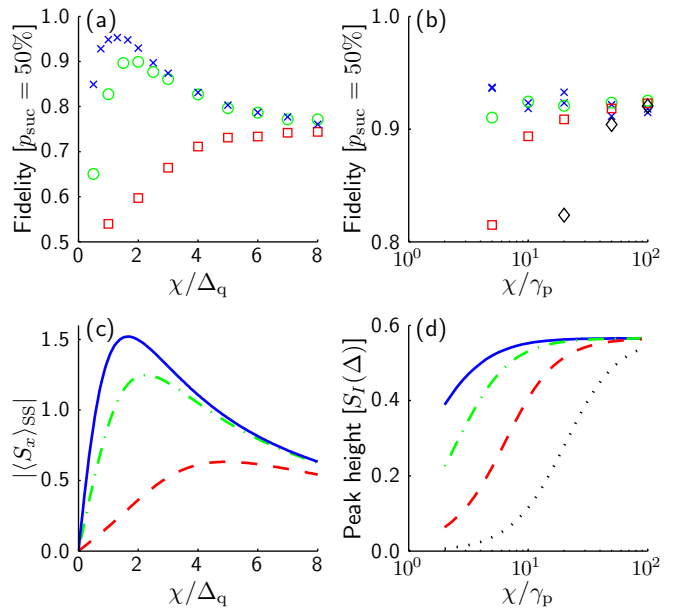}
  \caption{(Color online) Panels (a) and (b) show as a function of the
    resonant Rabi frequency, $\chi$, the obtained fidelity, $F$, with
    a success probability, $\psuccess = 50\%$. Panel (a): DC-analysis
    of $I(t)$ with $\theta = -\pi/2$, and $\Deltaa/\gammap$ given by
    $10$ (blue crosses), $1$ (green circles), and $0.3$ (red
    squares). Panel (b): AC-analysis of $I(t)$ with $\theta = 0$, and
    $\Deltaa/\gammap$ equal to $0$ (blue crosses), $1$ (green
    circles), $3$ (red squared), and $10$ (black diamonds). The
    simulations corresponding to the blue crosses have been performed
    twice in order to depict the statistical uncertainty of $F$. (c)
    The magnitude of the triplet-space steady-state value of $\S_x$ as
    given by Eq.~(\ref{eq:Mean_Sx_triplet_SS}) for $\Deltaa/\gammap =
    10$ (blue-solid line), $1$ (green, dash-dotted line), and $0.3$
    (red-dashed line). (d) The height of the spectral peak in
    $S_I(\Delta)$ when $\Deltaa/\gammap = 0$ (blue-solid line), $1$
    (green-dash-dotted line), $3$ (red-dashed line), and $10$
    (black-dotted line).}
  \label{fig:CompSimResults}
\end{figure}

Following the strategy presented in Sec.~\ref{sec:Practical-strategy},
we consider first the use of the photo-current mean value signal from
Eq.~(\ref{eq:zetameanI}) for distinguishing between the singlet state
and the triplet space. According to Eq.~(\ref{eq:I(t)_ad_elim}) it is
feasible to choose the local oscillator phase, $\theta =
-\frac{\pi}{2}$, such that $I(t) = \sqrt{\gammap\etaeff}\mean{\S_x} +
\xi(t)$ does not contain a background contribution from the cavity
field (note $\theta_{\kappa} \approx 0$). The ensemble average of
Eq.~(\ref{eq:zetameanI}) then becomes:
\begin{equation}
\label{eq:Ensemble_mean_zetameanI}
  \begin{split}
  \mean{\zetameanI}_{\mathrm{E}} & =\sqrt{\frac{\gammap\etaeff}{T}}
    \int_0^T\mean{\mean{\S_x(t)}}_{\mathrm{E}}dt \\ &\approx
    \left\{
    \begin{matrix}
      \sqrt{\gammap T\etaeff}\mean{\S_x}_{\mathrm{SS}} & \text{(triplet)} \\
      0 & \text{(singlet)}
    \end{matrix}
    \right.,
  \end{split}
\end{equation}
where the triplet-space steady-state value of $\S_x$ is predicted to
be (details given in Appendix~\ref{app:steady-state-prop_triplet}):
\begin{equation}
\label{eq:Mean_Sx_triplet_SS}
 \mean{\S_x}_{\mathrm{SS}} = \frac{-2\chi\Deltaa(\gammap^2 + 4\Deltaa^2 + 2\chi^2)}
 {(\gammap^2 + 4\Deltaa^2)(\gammap^2 + \Deltaa^2 + \chi^2) + \frac{3}{4}\chi^4},
\end{equation}
and the approximation assumes that $\mean{S_x(t)}$ corresponds to
$\mean{\S_x}_{\mathrm{SS}}$ most of the time, i.e.~$\gammap T \gg
1$. For the singlet state the $\delta$-correlated nature of $\xi(t)$
leads to the variance, $\mathrm{Var}(\zetameanI) = 1$, while the
distribution is broader for the triplet space due to temporal
variations in $\mean{\S_x}$ (see Fig.~\ref{fig:DCSimulationExample}(b)
and the discussion below). Despite the crudeness of the approximation
in Eq.~(\ref{eq:Ensemble_mean_zetameanI}) it is clear from the above
discussion that an effective distinction between the singlet- and
triplet-spaces is possible when $\gammap T \etaeff
\mean{\S_x}^2_{\mathrm{SS}} \gg 1$, and that the parameters, $\Deltaa$
and $\chi$, should be optimized in order to maximize
$\mean{\S_x}_{\mathrm{SS}}$ according to
Eq.~(\ref{eq:Mean_Sx_triplet_SS}). When $\Deltaa \gg \gammap$, the
maximum value of Eq.~(\ref{eq:Mean_Sx_triplet_SS}) is
$\mean{\S_x}_{\mathrm{SS}} \approx 1.52$ obtained when $\chi \approx
1.65\Deltaa$ (see also Fig.~\ref{fig:CompSimResults}(c)).

Now, consider Fig.~\ref{fig:DCSimulationExample} exemplifying the
entanglement-generating process. In panel (a) the integrated
photo-current, $\zetameanI$, of Eq.~(\ref{eq:zetameanI}) has been
plotted for the two individual simulation runs, which were already
discussed in Fig.~\ref{fig:BayesianExamples}(a). As time evolves, the
black-circled curve shows an increasing value of $|\zetameanI|$, which
reflects the fact that the photo-current, $I(t)$ (black circles), in
the upper part of Fig.~\ref{fig:BayesianExamples}(a) has a mean value
slightly below zero as a consequence of $\mean{\S_x}$ being non-zero
for a triplet-state simulation instance. In contrast, in
Fig.~\ref{fig:BayesianExamples}(a) the photo-current (blue crosses)
representing a simulation instance ending up in the singlet state is
closer to zero on average, which again is reflected in the measurement
signal (blue crosses) in Fig.~\ref{fig:DCSimulationExample}(a). Now,
the practical acceptance criterion consists simply of keeping a given
state $\dens$ provided that $\zetameanI$ ends up at time $\gammap T =
10$ between the red-dotted lines in
Fig.~\ref{fig:DCSimulationExample}(a), i.e.~if $|\zetameanI| \le
\zetameanIthr$ for a pre-selected value of $\zetameanIthr$.

Using the entire set of simulations, the distribution of $\zetameanI$
has been plotted in panel (b) showing a clear double-peak
structure. By distinguishing between high ($\ge 0.8$) and low ($\le
0.2$) overlap with $\ket{-}$, we clearly see that each peak
corresponds to either the singlet or triplet space. The solid lines
are Gaussian functions with unit variance (the shot-noise level of the
homodyne-detection procedure) and mean values predicted by the crude
approximation of Eq.~(\ref{eq:Ensemble_mean_zetameanI}). The
singlet-state (crosses) is modeled accurately since $\mean{\S_x} = 0$
is exact and the only variation arises from the random shot noise of
the measurement. However, for the triplet state the simulated
distribution is evidently broader and asymmetric --- the additional
width arises from the qubit dynamic evolution within the triplet-state
manifold leading to a variation in $\S_x$. Again, the red-dotted lines
depict the acceptance window, which clearly selects most of the
singlet-state events; however, a small fraction of the undesired
triplet-state occurrences are also included. This effect illustrates
the fact that the experimentally simple acceptance criterion,
$|\zetameanI| \le \zetameanI^{\mathrm{thr}}$, is less accurate than
the complete calculation of $\dens$ discussed in
Sec.~\ref{sec:Optimal_strategy}. In fact, if the optimal method of
accepting states with $\mean{-|\dens|-} \ge \Fmin$ for some selected
value of $\Fmin$ is used the undesired instances from the triplet
space with poor singlet-state overlap would simply not occur.

The value of $\psuccess$ and $F$ can be calculated as a function of
$\zetameanI^{\mathrm{thr}}$, i.e.~for various widths of the acceptance
window, as shown in panel (c). Clearly, for small, increasing values
of $\zetameanI^{\mathrm{thr}}$ the success probability grows quickly
without much degradation in fidelity since the acceptance window
selects predominantly the states with a high singlet-state
overlap. When the 50-percent success probability is reached, a further
increase of $\zetameanI^{\mathrm{thr}}$ must incorporate some
triplet-state instances with a loss of fidelity as a result. These
observations can also be shown as an $F$-versus-$\psuccess$ plot, see
panel (d). Here the ultimate limit (shown by a blue-dashed line) can
be obtained if the singlet- and triplet-spaces present distribution
functions like those of panel (b) but being entirely separate.

The fidelity obtained at 50\% success probability is shown in
Fig.~\ref{fig:CompSimResults}(a) for various values of $\Deltaa$ and
$\chi$. The variation in this fidelity can then be compared to the
triplet-state steady-state mean value, $\mean{\S_x}_{\mathrm{SS}}$,
which has been plotted in Fig.~\ref{fig:CompSimResults}(c) for the
same parameter settings of $\Deltaa$ and $\chi$. The correlation
between these figures is evident, which confirms the simple picture
discussed around Eq.~(\ref{eq:Ensemble_mean_zetameanI}) that the
triplet-state imprint onto the photo-current must be maximized for
optimizing the performance of the protocol.

\subsubsection{AC-analysis of the photo-current}
\label{sec:ac-analysis}
\begin{figure}
  \centering
  \includegraphics[width=\linewidth]{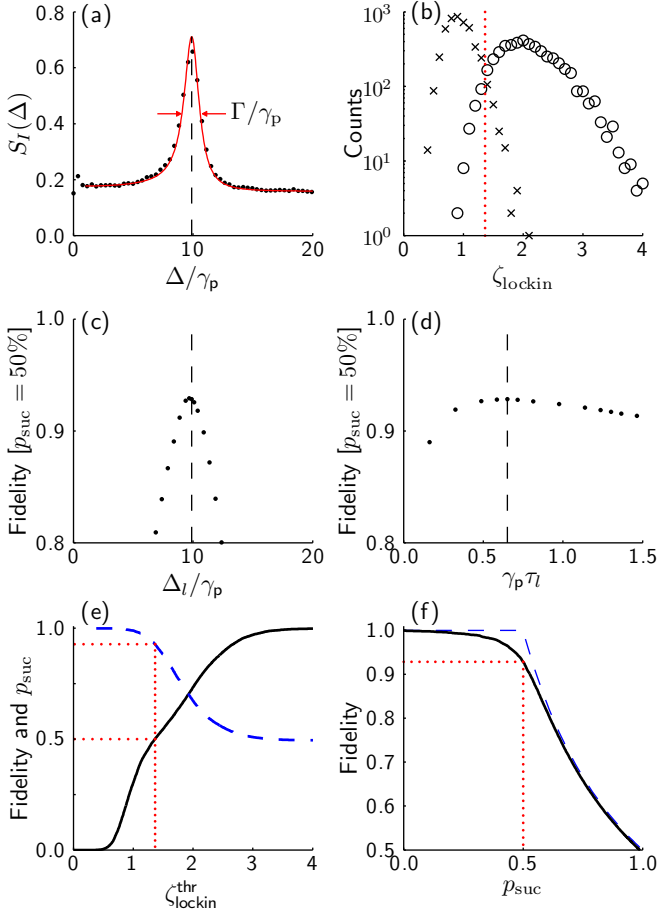}
  \caption{(Color online) All graphs are based on 10,000 simulations
    with $\chi = 10\gammap$, $\Deltaa = 0$, $\gammap T = 10$. (a) The
    simulated spectrum (dots) using the periodogram, $P_I(\Delta)$
    [Eq.~(\ref{eq:def_periodogram})] on simulation instances with
    $\mean{-|\dens|-} \le 0.2$, compared to the analytical spectrum,
    $S_I(\Delta)$ (solid line), of Eq.~(\ref{eq:Def_S_I(Delta)}). The
    dashed line at $\Delta_0 = 9.89\gammap$ marks the maximum of
    $S_I(\Delta)$, and $\Gamma$ denotes the FWHM of $S_I(\Delta)$. (b)
    The distribution of $\zetalockin$ distinguished by a high ($\ge
    0.8$, crosses) and a low ($\le 0.2$, circles) singlet-state
    overlap. The lockin parameters, $\Omegalockin = \Delta_0$ and
    $\taulockin = \Gamma^{-1}$, follow the characteristics of
    $S_I(\Delta)$ from panel (a), and the red-dotted line marks an
    acceptance criterion, $\zetalockin \le
    \zetalockin^{\mathrm{thr}}$, with $\psuccess = 50\%$. Panels (c)
    and (d) show the obtained fidelity (with $\psuccess = 50\%$) for
    varying lockin parameters, $\Omegalockin$ and $\taulockin$,
    respectively. The dashed lines mark the parameters used in panel
    (b). Panels (e) and (f) show the variations in fidelity and
    success probability for varying acceptance thresholds. The
    red-dotted lines mark the obtained fidelity $F = 93\%$ at
    $\psuccess = 50\%$.}
  \label{fig:LockinCharacteristics}
\end{figure}
The spectrum of the photo-current is defined as:
\begin{equation}
\label{eq:Def_S_I(Delta)}
  S_I(\Delta) = \frac{1}{2\pi}\int_{-\infty}^{\infty}R(t;\tau)e^{i\Delta\tau}d\tau,
\end{equation}
which generally depends on the time, $t$, in the initial transient
regime but is time-independent in steady state. For the singlet state
the spectrum is flat, $S_I(\Delta) = \frac{1}{2\pi}$, corresponding to
the shot noise level of the homodyne detection apparatus. The triplet
space is distinguished from the singlet state by identifying a
spectral peak in this flat background. In similarity with the
expectations discussed around Eq.~(\ref{eq:Ensemble_mean_zetameanI})
for the DC-analysis, we shall here use the steady-state spectrum of
the triplet space to predict the optimization of $\Deltaa$, $\chi$,
$\Omegalockin$, and $\taulockin$ for best performance of the
measurement signal, $\zetalockin$. As a first step, consider the
dynamical mean-value equations for $\vec{S} = [\S_x, \S_y,
\S_z]^{\mathrm{T}}$, which follow immediately from
Eqs.~(\ref{eq:ddt_Splus_triplet_noDecay})-(\ref{eq:ddt_Sz_triplet_noDecay})
with $\chi$ being real:
\begin{equation}
  \frac{\partial\mean{\vec{S}}}{\partial t} = \vec{Q}\times\mean{\vec{S}}
     + \gammap[\ldots],
\end{equation}
where the coherent driving vector is given by $\vec{Q} = [\chi, 0,
\Deltaa]^{\mathrm{T}}$, and the $\gammap$-term (with quadratic
$\vec{S}$-components left out for clarity) tends to drive
$\mean{\vec{S}}$ toward the vector $[0, 0, -2]$, i.e., the state
$\ket{\mathrm{gg}}$. The modulation of the photo-current, through
$\mean{\S_x}$ and $\mean{\S_y}$ according to
Eq.~(\ref{eq:I(t)_ad_elim}), is largest when the spin vector is
allowed to sweep across the full sphere, i.e.~we expect that $\chi \gg
\gammap, \Deltaa$ is a good choice in order to maintain a significant
level of excitation. We note from the expression of $\vec{Q}$ that
$\mean{\vec{S}}$ will primarily be spinning around the $x$-axis, which
leads to significant oscillations in $\mean{\S_y}$. For this reason,
the local-oscillator-phase choice, $\theta = 0$, is natural. The
oscillations in $I(t)$ are then superposed on a constant background
level $\propto \Re\{\alphac\}$.

The spectrum of the simulated current, $I_j = I(t_j)$ where $j$ runs
over the discrete times separated by $dt$, can be conveniently
estimated by the periodogram
\begin{equation}
\label{eq:def_periodogram}
  P_I(\Delta) = \frac{dt}{2\pi n}\left|\sum_{j=1}^n I_j e^{-i\Delta t_j}\right|^2,
\end{equation}
which is essentially the modulus square of the discrete Fourier
transform of $I(t)$. The front factor ensures the correct value,
$P_I(\Delta) = \frac{1}{2\pi}$, for the shot noise background, and we
subtract from $I(t)$ the constant contribution of the bare cavity
$\propto \Re\{\alphac\}$ prior to insertion into $P_I(\Delta)$.

Turning to the simulation, Fig.~\ref{fig:LockinCharacteristics}(a)
shows the simulated spectrum for the subset of instances, which
collapse into the triplet space, in comparison to the expectation from
Eq.~(\ref{eq:Def_S_I(Delta)}), which in steady state is given by
Eq.~(\ref{eq:Calculate_S_I(Delta)}). Since the integration time,
$\gammap T = 10$, is significantly larger than unity, this
steady-state expression does in fact match the simulated curve very
well. In the limit, $\chi \gg \Deltaa,\gammap$, numerical inspection
of Eq.~(\ref{eq:Calculate_S_I(Delta)}) reveals a Lorentzian peak
centered around the generalized Rabi frequency, $\Omega = \sqrt{\chi^2
  + \Deltaa^2}$, and with full-width at half maximum (FWHM) $\Gamma =
\frac{3}{2}\gammap$ (for the solid curve in
Fig.~\ref{fig:LockinCharacteristics}(a), the maximum is placed at
$\Delta_0 = 0.989\Omega$ with $\Gamma = 1.54\gammap$). In order to
distinguish between the singlet- and triplet-space part of the initial
state, $\ket{\psi}$, we must establish the absence or presence of this
spectral peak in each individual simulation run. To this end, the
lockin parameters of Eq.~(\ref{eq:zetalockin}) are chosen as
$\Omegalockin = \Delta_0$ and $\taulockin = \Gamma^{-1}$, and as can
be seen from Fig.~\ref{fig:LockinCharacteristics}(b), the measurement
signal, $\zetalockin$, is indeed capable of separating the singlet
state from the triplet space. The robustness of this procedure to
errors in the lockin parameters is depicted in
Fig.~\ref{fig:LockinCharacteristics}(c,d), which show that
$\Omegalockin$ must obviously match the position of the spectral peak
with an accuracy set by $\Gamma$ and that $\taulockin \approx
\Gamma^{-1}$ provides the best match to the bandwidth of the signal
peak. The obtained fidelity and success probability while varying the
acceptance criterion, $\zetalockin \le \zetalockin^{\mathrm{thr}}$,
can be seen in Fig.~\ref{fig:LockinCharacteristics}(e,f). These graphs
are quite similar to the corresponding results for the DC-measurements
in Fig.~\ref{fig:DCSimulationExample}(c,d); however, if a low success
probability is accepted, the obtained fidelity seems to be better.

The optimization of the qubit-driving parameters, $\chi$ and
$\Deltaa$, are examined in Fig.~\ref{fig:CompSimResults}(b,d). Panel
(b) shows the obtained fidelity in various simulations runs, and there
is a clear correlation with the calculated spectral-peak height of
$S_I(\Delta)$ shown in panel (d). In similarity with the DC-analysis
in Sec.~\ref{sec:dc-analysis}, the present AC-analysis of $I(t)$ is
optimized in terms of $\chi$ and $\Deltaa$ simply by maximizing the
triplet-state steady-state spectral peak height, and
Fig.~\ref{fig:CompSimResults}(d) presents the practical condition,
$\chi \gtrsim 10\cdot\max(\gammap,\Deltaa)$, for this optimization.

\subsection{Practical versus optimal extraction of information}
\label{sec:Prac_vs_opt_info}
\begin{figure}
  \centering
  \includegraphics[width=\linewidth]{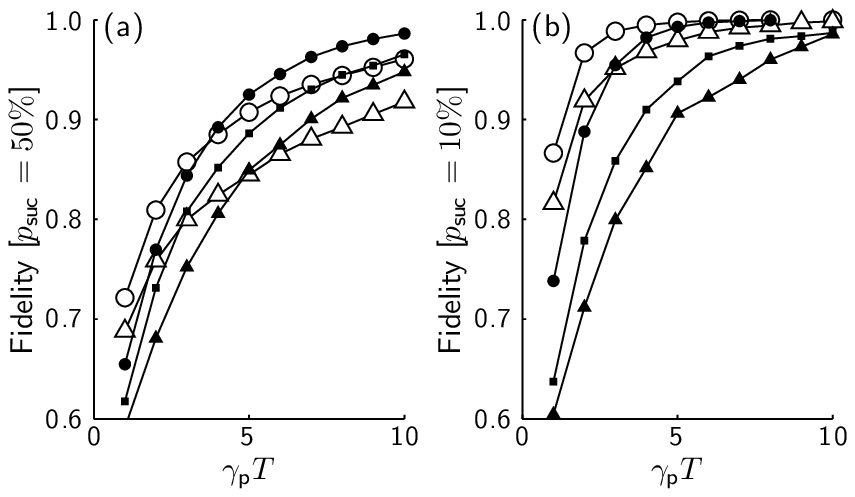}
  \caption{The obtained fidelity versus integration time when
    $\psuccess = 50\%$ (panel a) or $\psuccess = 10\%$ (panel
    b). Fidelities obtained by optimal analysis based on
    $\mean{-|\dens|-} \ge \Fmin$ are shown for $\theta = 0$ (open
    circles) and $\theta = -\pi/2$ (solid circles). The lockin
    analysis, $\zetalockin \le \zetalockin^{\mathrm{thr}}$ from
    Sec.~\ref{sec:ac-analysis}, gives rise to the open triangles,
    while the integrated-photo-current analysis, $|\zetameanI| \le
    \zetameanI^{\mathrm{thr}}$ from Sec.~\ref{sec:dc-analysis}, leads
    to the solid triangles. The solid squares represent the displaced,
    time-weighted measurement signal described in the text.}
  \label{fig:ACDCcomparison}
\end{figure}
In Fig.~\ref{fig:ACDCcomparison} the performance of the DC-analysis of
$I(t)$ as described in Sec.~\ref{sec:dc-analysis} (solid triangles)
can be directly compared to an optimal extraction of information from
the full $I(t)$ (solid circles) given the local-oscillator phase,
$\theta = -\pi/2$. Likewise, the lockin-based AC-analysis of $I(t)$
[Sec.~\ref{sec:ac-analysis}] shown with open triangles can be related
to an optimal information extraction (open circles) given the phase
choice, $\theta = 0$. In comparison to the optimal extraction of
information, the simple and more robust approaches require
approximately twice the time for obtaining a given fidelity with a
given success rate.

For the DC-analysis protocol the detection record during the initial
transient dynamics of duration $\approx \gammap^{-1}$ does not bear
much information since neither the singlet- nor triplet-space part of
$\ket{\psi}$ gives rise to a non-zero value of $\mean{\vec{S}}$ in the
initial time range, $0 < t \lesssim \gammap^{-1}$, in which the two
qubit spins are pointing in opposite directions. Hence, by weighting
the integral in Eq.~(\ref{eq:zetameanI}) by the function
$(1-e^{-\gammap t})$, we do not loose information but a smaller amount
of shot noise is accumulated in this transient part of the
protocol. Combining this weighting procedure with an acceptance window
shifted by 0.5 toward the right in
Fig.~\ref{fig:DCSimulationExample}(b), we obtain the improved
fidelities shown by solid squares in Fig.~\ref{fig:ACDCcomparison},
and the DC-analysis protocol narrows in on the full calculation of
$\dens$.

The AC-analysis protocol is based on the correlation
function~(\ref{eq:R(t;tau)_ad_elim}), which contains quadratic moments
of $\vec{S}$ and hence is able to deliver an oscillatory signal
starting already from $t=0$. Considering
Fig.~\ref{fig:ACDCcomparison}, we ascribe this fact as the reason for
the slightly better performance of spin-precession-based protocols
(open symbols) in comparison to the spin-mean-value-based protocols
(solid symbols) at short integration times.

\section{The influence of decoherence processes}
\label{sec:Decoherence_processes}
\begin{figure}
  \centering
  \includegraphics[width=\linewidth]{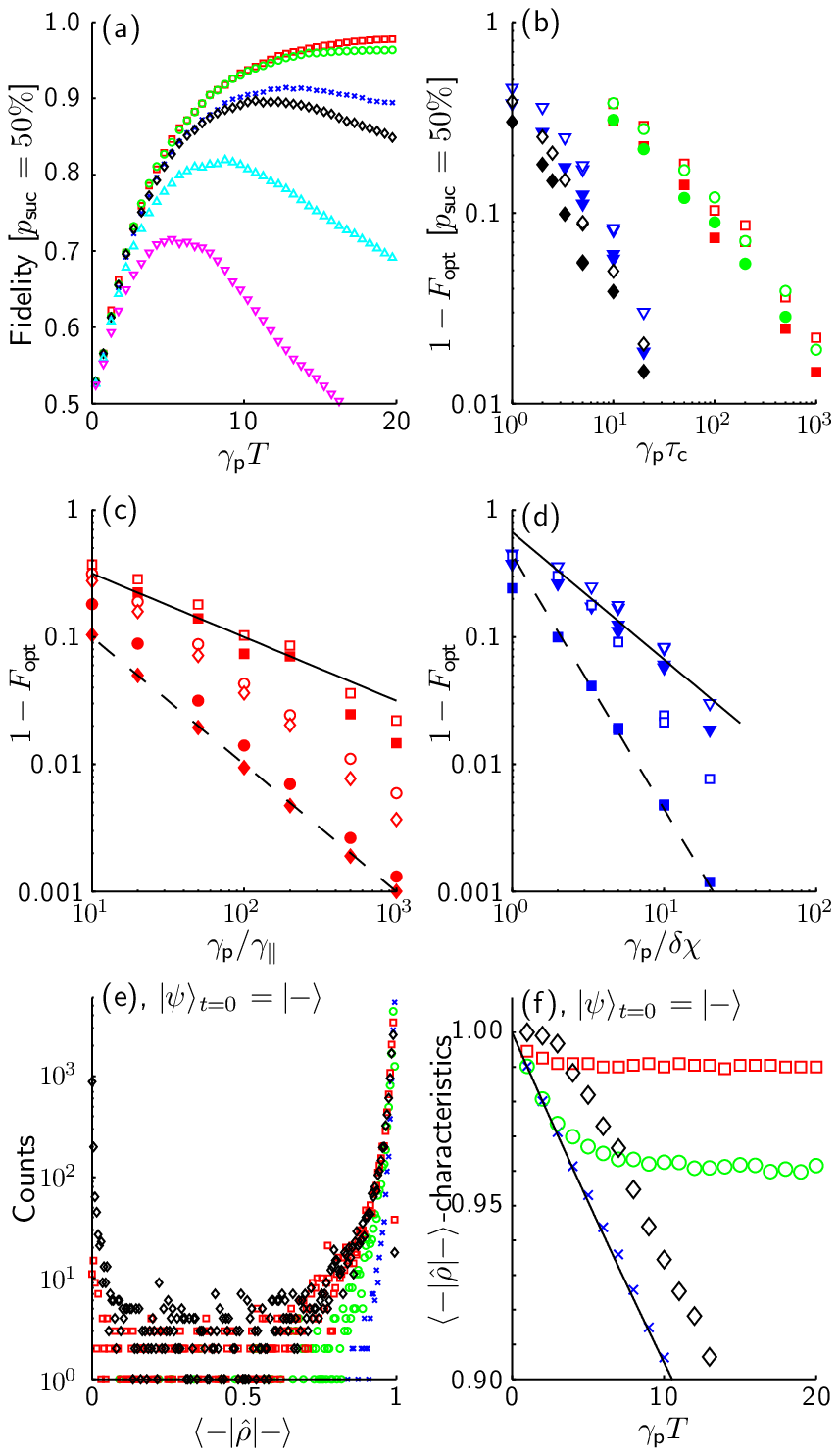}
  \caption{(Color online) All panels show simulation results for
    $\Deltaa = 10\gammap$, $\chi = 16.5\gammap$, and $\theta =
    -\pi/2$. (a) The obtained fidelity when $\psuccess = 50\%$ using
    $|\zetameanI| \le \zetameanI^{\mathrm{thr}}$ while varying
    $\gammapar/\gammap = 1\times 10^{-3}$ (Red squares), $2\times
    10^{-3}$ (green circles), $5\times 10^{-3}$ (blue crosses),
    $1\times 10^{-2}$ (black diamonds), $2\times 10^{-2}$ (cyan tip-up
    triangles), $5\times 10^{-2}$ (magenta tip-down triangles). (b)
    The optimum fidelity when $\psuccess = 50\%$ for various
    characteristic decoherence times $\tauc$. Red squares: Population
    decay with $\tauc = \gammapar^{-1}$, green circles: Phase
    decoherence with $\tauc = \tau$, blue triangles: Inhomogeneous
    coupling strength with $\tauc = \delta\chi$, black diamonds:
    Inhomogeneous qubit frequency with $\tauc = \delta\omegaa$.  Open
    and closed symbols are obtained using $|\zetameanI| \le
    \zetameanI^{\mathrm{thr}}$ and $\mean{-|\dens|-} \ge \Fmin$,
    respectively (this holds also for panels c and d). (c) Fidelity
    versus $\gammapar^{-1}$ for $\psuccess = 50\%$ (squares),
    $\psuccess = 30\%$ (circles), and $\psuccess = 10\%$
    (diamonds). Solid and dashed lines denote $\Fopt =
    1-\sqrt{\gammapar/\gammap}$ and $\Fopt = 1 -\gammapar/\gammap$,
    respectively. (d) Fidelity versus $\delta\chi^{-1}$ for $\psuccess
    = 50\%$ (triangles) and $\psuccess = 10\%$ (squares). Solid and
    dashed lines correspond to slopes of $-1$ and $-2$,
    respectively. In panels (e) and (f) $\ket{\psi} = \ket{-}$ at
    $t=0$ and $\gammapar/\gammap = 0.01$. (e) singlet-state overlap at
    $\gammap T = 1$ (blue crosses), $2$ (green circles), $5$ (red
    squares), and $20$ (black diamonds). (f) average value of
    $\langle-|\dens|-\rangle$ for all incidences (blue crosses) and
    for those with $\langle-|\dens|-\rangle \ge 0.5$ (green
    circles). Red squares: Most probable value of
    $\langle-|\dens|-\rangle$. Black diamonds: Fraction of states with
    $\langle-|\dens|-\rangle \ge 0.5$. Solid line: $\exp(-\gammapar
    T)$.}
  \label{fig:DecayData}
\end{figure}
This section estimates the effect of decoherence processes on the
obtainable fidelity. We note that if such processes are strong, the
optimum parameter settings as exemplified by
Fig.~\ref{fig:CompSimResults} might change. Instead of performing a
full-scale analysis of such possible changes, we simply add
decoherence processes but keep the measurement protocols and
acceptance criteria. The analysis presents a lower bound of the
obtainable fidelities an must be a good approximation in the limit of
high fidelities. Only the local-oscillator-phase choice of $\theta =
-\pi/2$ relevant for the DC-analysis protocol is discussed below ---
the case of $\theta = 0$ presents similar features.

To exemplify the simulation procedure, a qubit-population decay is
introduced in Fig.~\ref{fig:DecayData}(a) with varying values of
$\gammapar$ (modeled by the decay operators $\c_2$ and $\c_3$ in
Sec.~\ref{sec:stochastic_ME}). Since the decay process deteriorates
the desired singlet state in the long-integration-time limit there
exists an optimum integration time and a corresponding optimum
fidelity. This fidelity has been plotted in
Fig.~\ref{fig:DecayData}(b) --- see the figure caption for simulation
details. In a similar manner, the effect of qubit phase decay can be
modeled by assigning a finite value to $\tau$ in the operators $\c_4$
and $\c_5$. Furthermore, one may consider the case that the two qubits
are not coupled in the exact same way to the cavity. A small
inhomogeneity in the coupling strength for each qubit, $g_1 = g +
\frac{\delta g}{2}$ and $g_2 = g - \frac{\delta g}{2}$, will lead to a
slight difference in resonant Rabi frequency, $\delta\chi =
2\alphac\delta g$, or alternatively, a small difference in qubit
resonance frequency, $\delta\omegaa$, could be present. The effect of
these non-ideal scenarios are compared in Fig.~\ref{fig:DecayData}(b)
showing that qubit population decay (red squares) and dephasing (green
circles) behave in approximately the same way, while the
inhomogeneities in Rabi frequency (blue triangles) or qubit detuning
(black diamonds) follow their own distinct trend. These observations
can be complemented by the equations of motion for the singlet-state
population (only the deterministic part, i.e.~take $\eta = 0$):
\begin{equation}
\label{eq:ddt_rho_minusminus_determ}
  \begin{split}
      \frac{d\rho_{-,-}}{dt} = &-\left(\frac{1}{\tau}+\gammapar\right)
     \rho_{-,-} + \frac{1}{\tau}\rho_{+,+} + \gammapar\rho_{\mathrm{ee,ee}} \\
     &-\frac{i\delta\omegaa}{2}(\rho_{+,-} - \rho_{-,+}) \\
    &-\frac{i\delta\chi}{2}\left(\frac{\rho_{\mathrm{gg,-}}-\rho_{\mathrm{ee,-}}}
    {\sqrt{2}}  - \frac{\rho_{\mathrm{-,ee}}-\rho_{\mathrm{-,gg}}}{\sqrt{2}}\right)
  \end{split}
\end{equation}
Evidently, the two rates $\gammapar$ and $\tau^{-1}$ enter on the same
footing and are responsible for the de-population of the singlet
state. In the high-fidelity limit ($\rho_{-,-} \approx 1$,
$\rho_{+,+}, \rho_{\mathrm{ee,ee}} \approx 0$) one would expect the
infidelity, $1-F$, to increase linearly with these rates. Likewise,
the inhomogeneities parametrized by $\delta\omegaa$ and $\delta\chi$
seem to be comparable in effect --- they attempt to drive coherently
the population from the singlet state toward $\ket{+}$ (the
$\delta\omegaa$-term) or $\frac{1}{\sqrt{2}}(\ket{\mathrm{gg}} -
\ket{\mathrm{ee}})$ (the $\delta\chi$-term). In the high-fidelity
limit the coherence terms, $\rho_{+,-}$, $\rho_{\mathrm{gg},-}$ and
$\rho_{\mathrm{ee},-}$, must be polarized before the singlet-state
population can be driven, and hence the infidelity is expected to
increase quadratically with $\delta\chi$ or $\delta\omegaa$. In order
to exemplify these scaling behaviors, consider the red squares and
blue triangles of Fig.~\ref{fig:DecayData}(b), which have been
re-plotted in panels (c) and (d), respectively. These data scale
roughly as the solid lines, which in the double-logarithmic plots have
slopes $-\frac{1}{2}$ and $-1$ in panels (c) and (d),
respectively. This does not correspond to the scaling behavior
discussed above; however, by accepting a smaller success probability,
the fidelity increases and the expected scaling is found in the
high-fidelity limit. In fact, for the best case shown in panel (c)
with optimal information extraction and $\psuccess = 10\%$ (solid red
diamonds), the infidelity becomes, $1-\Fopt \approx
\gammapar/\gammap$. At first glance this is surprising since the
integration time for obtaining high fidelities is typically exceeding
$10\gammap^{-1}$ as exemplified in panel (a); however, the
infidelity-penalty is not exceeding $10\gammapar/\gammap$ but is equal
to roughly one unit of $\gammapar/\gammap$. A thorough examination of
the continuous measurement process in the high-fidelity limit is
required to explain this observation: Consider panels (e,f) based on
10,000 simulations in presence of qubit-population decay and using the
singlet-state as the initial state, i.e.~$\mean{-|\dens|-} = 1$ at
$t=0$. Panel (e) shows the distribution of $\mean{-|\dens|-}$ for
various times, which is seen to increase in width during the first few
$\gammap T$ but then settles to an almost constant distribution
(compare red squares and black diamonds). However, a sharp feature is
emerging around $\mean{-|\dens|-} = 0$ showing an increasing
population within the triplet space. Now, the ensemble mean value of
$\mean{-|\dens|-}$ based on these distributions is calculated and
plotted in panel (f) (blue crosses). The dynamics of this ensemble
mean value is governed simply by the deterministic part of the master
equation~(\ref{eq:MasterEq_original}) (since
$\mean{\xi(t)}_{\mathrm{E}} = 0$) which for the $\rho_{-,-}$-component
is given by Eq.~(\ref{eq:ddt_rho_minusminus_determ}). In the
high-fidelity limit the presence of population decay leads to
$\frac{\partial\rho_{-,-}}{\partial t} \approx -\gammapar\rho_{-,-}$,
the solution of which is the solid line in panel (f) confirming the
simulations. Nonetheless, the distribution is strongly peaked around
zero and unity. In comparison, a drastically different sub-ensemble
mean value of $\mean{-|\dens|-}$ can be obtained if we are able to
select the best singlet-state candidates. For instance, by
conditioning the sub-ensemble on $\mean{-|\dens|-} \ge 0.5$, the green
circles are obtained in panel (f). Furthermore, the most probable
value of $\mean{-|\dens|-}$ (i.e.~the maximum location of the
distributions in panel (e)) turns out as the red squares in panel
(f). Clearly, after a few times $\gammap T$ these conditioned
observables settle to a steady value. This does not contradict the
fact that the state is decaying --- the fraction of states with
$\mean{-|\dens|-} \ge 0.5$ is decreasing steadily as shown by black
diamonds in panel (f). To re-capitulate the above discussion: Once the
singlet state has been established the continuous measurement either
preserves it with a high fidelity or the state jumps into the triplet
space. Since the characteristic time for updating our knowledge is
$\gammap^{-1}$, the infidelity of the preserved singlet state is
approximately $\gammapar/\gammap$, which explains the steady value of
the red squares in panel (f) and the observed dashed-line scaling in
panel (c) (since exactly the very best states are selected in this
case). A similar set of arguments can be made for the high-fidelity
limit in panel (d) for the case of inhomogeneous cavity-qubit
coupling, and we note in both cases that the dashed-line slope is in
general maintained also for the DC-analysis-based acceptance
criterion, $|\zetameanI| \le \zetameanI^{\mathrm{thr}}$. We note that
by selecting the best states (in particular when using the optimal
extraction of information) the observed infidelity is determined by
the ability of the measurement to preserve the state and not sensitive
to the statistics of the finite number of simulations.

In the cases of qubit-population or phase decay, the product
$\gammap\tauc$ is essentially equal to (up to factors of two) the
cooperativity parameter $C = \frac{2g^2}{\kappa\gammaperp}$, where
$\gammaperp = \frac{1}{\tau} + \frac{\gammapar}{2}$. In the limit of
maximum success probability, $\psuccess = 0.5$, the above observations
conclude that $1-\Fopt \approx \frac{1}{\sqrt{C}}$, which is similar
to the figure of merit for deterministic protocols of cavity QED
\cite{Sorensen.PhysRevLett.91.097905(2003)}. The measurement process
seems to counteract effectively the loss of coherence inherent in the
bad-cavity limit, and by accepting a moderately lower success
probability, the states with an infidelity of $1-\Fopt \approx
\frac{1}{C}$ can be conditionally prepared. An infidelity scaling as
$\frac{1}{C}$ was also obtained in the heralded protocol of Ref.
\cite{Sorensen.PhysRevLett.91.097905(2003)}.

\section{Discussion}
\label{sec:discussion}
From a fundamental perspective, the correlated decay with rate
$\gammap$ is the key mechanism for the entanglement-generation
protocols. The original product state with opposite qubit-spins gives
rise to a 50\% overlap with $\ket{-}$, which allows the measurement
process to induce a collapse into the desired singlet state with a
high success probability. However, the information of the two-qubit
state must be transferred to the cavity field and subsequently leave
the cavity before reaching the homodyne-detection apparatus, and it is
exactly the decay rate, $\gammap$, which describes the combined rate
of this information flow. When the cavity field is adiabatically
eliminated the decay rate, $\gammap$, materializes explicitly as a
strength parameter in the homodyne-detection photo-current, see
e.g.~Eqs.~(\ref{eq:I(t)_ad_elim})
and~(\ref{eq:R(t;tau)_ad_elim}). Hence, in our analysis it does not
make sense to consider the particular case of $\gammap = 0$ (as was
done in Figs.~2 and 3 of Ref.~\cite{Hutchison.CanJPhys.87.225(2009)})
since no information is gained by the measurement.

Even though the measurement process is continuous, the discussion in
Sec.~\ref{sec:Decoherence_processes} revealed an effective jump-like
behavior of the quantum state, and we also remind that discreteness is
regained in the long-integration-time limit as exemplified clearly by
Fig.~\ref{fig:BayesianExamples}. These observations are not only of
fundamental importance --- the detection signal effectively monitors
any unwanted transitions into the triplet space. A simple feedback can
thus be implemented in order to establish and maintain a high
spin-singlet overlap in a continuous operating mode of the experiment.
We imagine that a continuous feedback strategy along the lines of
Ref.~\cite{Wang.PhysRevA.71.042309(2005)} can be developed taking into
account the known imprint of the triplet-state components onto the
photo-current.

The entanglement-generation protocol relies heavily on the fact that
the $\ket{-}$-state does not couple to the cavity field and the
interaction with the field effectively implements a QND measurement
\cite{Braginsky.RevModPhys.68.1(1996)} of the projection operator
$\hat{q} = \ket{-}\bra{-}$. This apparent QND character of the
protocol does not rely on the adiabatic elimination and the protocol
should be applicable outside the bad-cavity limit. However, the
optimization considerations of Sec.~\ref{sec:ent-generation-no-decay}
and the fidelity analysis of Sec.~\ref{sec:Decoherence_processes}
require the adiabatic approximation to be valid.
\section{Conclusion}
\label{sec:Conclusion}
A measurement-based entanglement-generating protocol has been
established for two qubits residing in a bad cavity.  The separation
of an initial state into either the singlet or triplet space
facilitates the establishment of entanglement, which is done optimally
by a stochastic-master-equation approach. In addition, two practical
methods have been discussed: (1) The use of the integrated
photo-current mean value and (2) a lock-in-based identification of
oscillations in the photo-current. The optimization of these methods
for best performance can be understood simply as maximizing the
imprint of the triplet-space part onto the photo-current. The
influence of qubit dissipation and inhomogeneities has been analyzed
such that obtainable fidelities can be estimated from relevant
experimental parameters.

\begin{acknowledgments}
  The authors acknowledge support from the EU integrated project AQUTE
  and the EU 7th Framework Programme collaborative project iQIT.
\end{acknowledgments}

\appendix

\section{Steady-state properties of the triplet space}
\label{app:steady-state-prop_triplet}
In this appendix the photo-current, $I(t)$, and its spectrum,
$S_I(\Delta)$, is calculated under the assumption that the two qubits
have reached steady state and also assuming that qubit-decay processes
are absent ($\gammapar = 0$, $\tau = \infty$). Under the latter
assumption the singlet and triplet spaces are decoupled and we
consider here only the three-dimensional triplet space. We shall also
ignore the information gain from the photo-current (i.e.~leaving out
the measurement-super-operator part of
Eq.~(\ref{eq:MasterEq_original})) in order to establish an a-priori
prediction of the photo-current.

Considering the master equation with the choices made above, the
dynamical equations for mean values of $\S_+$, $\S_-$, and $\S_z$ read
(with $\chi$ real and neglecting the last, Stark-shift term of
Eq.~(\ref{eq:Heff})):
\begin{align}
\label{eq:ddt_Splus_triplet_noDecay}
  \frac{\partial\mean{\S_+}}{\partial t} &= i\Deltaa\mean{\S_+}
    -\frac{i\chi}{2}\mean{\S_z} + \frac{\gammap}{2}\mean{\S_+\S_z}, \\
\label{eq:ddt_Sminus_triplet_noDecay}
  \frac{\partial\mean{\S_-}}{\partial t} &= -i\Deltaa\mean{\S_-}
    +\frac{i\chi}{2}\mean{\S_z} + \frac{\gammap}{2}\mean{\S_z\S_-}, \\
\label{eq:ddt_Sz_triplet_noDecay}
  \frac{\partial\mean{\S_z}}{\partial t} &= -i\chi[\mean{\S_+}-\mean{\S_-}]
    -2\gammap\mean{\S_+\S_-},
   \\ \notag &\rightarrow  -i\chi[\mean{\S_+}-\mean{\S_-}] +\gammap[
    \frac{1}{2}\mean{\S_z^2} - \mean{\S_z} - 4].
\end{align}
In Eq.~(\ref{eq:ddt_Sz_triplet_noDecay}) the second line is valid in
the special case of the triplet space since the two operators
$\hat{O}_1 = \S_+\S_-$ and $\hat{O}_2 = 2 + \frac{1}{2}\S_z -
\frac{1}{4}\S_z^2$ act identically on the triplet-state basis set:
$\hat{O}_j\ket{\mathrm{ee}} = 2\ket{\mathrm{ee}}$,
$\hat{O}_j\ket{\mathrm{+}} = 2\ket{\mathrm{+}}$,
$\hat{O}_j\ket{\mathrm{gg}} = 0$, for $j = 1,2$. We note from all
three equations above that the right-hand sides contain mean values of
the quadratic operators, $\S_+\S_z$, $\S_z\S_-$, and $\S_z^2$, and to
proceed the time-derivative of these mean values must be
calculated. In turn, cubic operators are introduced and the set of
equations seems endless. However, as exemplified above for
Eq.~(\ref{eq:ddt_Sz_triplet_noDecay}), by employing operator
identities valid in particular for the triplet space, the equations
become closed within an eight-dimensional space (corresponding to the
number of free parameters in the triplet-space density
matrix). Considering the column vector of mean-values: $\vec{x} =
[\mean{\S_+}, \mean{\S_-}, \mean{\S_z}, \mean{\S_+\S_z},
\mean{\S_z\S_-}, \mean{\S_+^2}, \mean{\S_-^2},
\mean{\S_z^2}]^{\mathrm{T}}$, the dynamical equations become after
some algebra:
\begin{equation}
  \frac{\partial\vec{x}}{\partial t} = \vec{A}\vec{x} - \vec{b},
\end{equation}
where the matrix, $\vec{A}$, and the column vector, $\vec{b}$, are
given by:
\begin{widetext}
  \begin{equation}
  \vec{A} =
  \begin{bmatrix}
    i\Deltaa & 0 & -\frac{i\chi}{2} & \frac{\gammap}{2} & 0 & 0 & 0 & 0 \\
    0 & -i\Deltaa & \frac{i\chi}{2} & 0 & \frac{\gammap}{2} & 0 & 0 & 0 \\
    -i\chi & i\chi & -\gammap & 0 & 0 & 0 & 0 & \frac{\gammap}{2} \\
   -4\gammap & 0 & \frac{i\chi}{2} & -3\gammap+i\Deltaa & 0 & -i\chi & 0 & 
      -\frac{3i\chi}{4} \\
   0 & -4\gammap & -\frac{i\chi}{2} & 0 & -3\gammap-i\Deltaa & 0 & i\chi &
      \frac{3i\chi}{4} \\
   -i\chi & 0 & 0 & -i\chi & 0 & -\gammap+2i\Deltaa & 0 & 0 \\
   0 & i\chi & 0 & 0 & i\chi & 0 & -\gammap-2i\Deltaa & 0 \\
   -2i\chi & 2i\chi & -2\gammap & -2i\chi & 2i\chi & 0 & 0 & -3\gammap
  \end{bmatrix},
  \quad
  \vec{b} =
  \begin{bmatrix}
    0 \\ 0 \\ 4\gammap \\ -2i\chi \\ 2i\chi \\ 0 \\ 0 \\ -8\gammap
  \end{bmatrix}.
\end{equation}
\end{widetext}
The steady-state value of $\vec{x}$ is now simply given by
$\vec{x}_{\mathrm{SS}} = \vec{A}^{-1}\vec{b}$, and since $\S_x = \S_+
+ \S_-$ the result of Eq.~(\ref{eq:Mean_Sx_triplet_SS}) follows after
some algebra.

Turning to the spectrum, $S_I(\Delta)$, of the photo-current as
defined in Eq.~(\ref{eq:Def_S_I(Delta)}), we note that correlation
functions between $\S_+$ and $\S_-$ are required according to
Eq.~(\ref{eq:R(t;tau)_ad_elim}). To this end, define first a new
vector:
\begin{equation}
  \vec{y}(t,\tau) =
  \begin{bmatrix}
    \mean{\S_+(t+\tau)\S_-(t)} \\ \mean{\S_-(t+\tau)\S_-(t)} \\ 
    \mean{\S_z(t+\tau)\S_-(t)} \\ \mean{\S_+(t+\tau)\S_z(t+\tau)\S_-(t)} \\ 
    \mean{\S_z(t+\tau)\S_-(t+\tau)\S_-(t)} \\ \mean{\S_+^2(t+\tau)\S_-(t)} \\ 
    \mean{\S_-^2(t+\tau)\S_-(t)} \\ \mean{\S_z^2(t+\tau)\S_-(t)}    
  \end{bmatrix},
\end{equation}
%
%
i.e.~this is the vector $\vec{x}$ evaluated at $t+\tau$ and multiplied
by $\S_-(t)$ \emph{inside} the mean value brackets
$\langle\ldots\rangle$. According to the quantum regression theorem,
the time-evolution of $\vec{y}(t,\tau)$ for $\tau \ge 0$ follows the
exact same equation as $\vec{x}(t)$,
i.e.~$\frac{\partial\vec{y}(t,\tau)}{\partial\tau} =
\vec{A}\vec{y}(t,\tau) - \vec{b}\mean{\S_-(t)}$, which has the solution:
\begin{equation}
\label{eq:Solution_y(tau)-y(inf)}
  [\vec{y}(t,\tau)-\vec{y}(t,\infty)] = e^{\vec{A}\tau}
    [\vec{y}(t,0)-\vec{y}(t,\infty)],
\end{equation}
where $\vec{y}(t,\infty) = \vec{A}^{-1}\vec{b}\mean{\S_-(t)} =
\vec{x}_{\mathrm{SS}}\mean{\S_-(t)}$. The first two entries of
$[\vec{y}(t,\tau)-\vec{y}(t,\infty)]$ are equal to
$\mean{\S_+(t+\tau),\S_-(t)}$ and $\mean{\S_-(t+\tau),\S_-(t)}$,
respectively, which (together with their complex conjugates) are
exactly the terms required in Eq.~(\ref{eq:R(t;tau)_ad_elim}). Since
in steady state, $R(t;-\tau) = R(t;\tau)$, the spectrum can be
calculated conveniently as an integral over $\tau \ge 0$:
\begin{widetext}
  \begin{equation}
  \label{eq:Calculate_S_I(Delta)}
  \begin{split}
  S(\Delta) &= \frac{1}{\pi}\int_0^{\infty}R(t;\tau)\cos(\Delta\tau)d\tau \\
  &= \frac{1}{2\pi} 
    +\frac{\gammap\etaeff}{\pi}\left\{ 
    \vec{v}^{\mathrm{T}}
    \int_0^{\infty}[\vec{y}(t,\tau)-\vec{y}(t,\infty)]\cos\Delta\tau d\tau
    + \mathrm{c.c.}\right\} \\
  &= \frac{1}{2\pi} 
    -\frac{\gammap\etaeff}{2\pi}\left\{
    \vec{v}^{\mathrm{T}}[(\vec{A}+i\Delta)^{-1}+(\vec{A}-i\Delta)^{-1}]
    [\vec{y}(t,0)-\vec{y}(t,\infty)] + \mathrm{c.c.}\right\} \\
  &= \frac{1}{2\pi} 
    -\frac{\gammap\etaeff}{2\pi}\left\{
    \vec{v}^{\mathrm{T}}[(\vec{A}+i\Delta)^{-1}+(\vec{A}-i\Delta)^{-1}]
    [(\vec{C}-\mean{\S_-})\vec{A}^{-1}\vec{b}+\vec{d}] + \mathrm{c.c.}\right\},
  \end{split}
\end{equation}
with $\vec{C}$, $\vec{d}$, and $\vec{v}$ defined as
\begin{equation}
  \label{eq:Def_C_d_v}
  \vec{C} =
  \begin{bmatrix}
    0 & 0 & \frac{1}{2} & 0 & 0 & 0 & 0 & -\frac{1}{4} \\
    0 & 0 & 0 & 0 & 0 & 0 & 1 & 0 \\
    0 & 0 & 0 & 0 & 1 & 0 & 0 & 0 \\
    0 & 0 & 0 & 0 & 0 & 0 & 0 & 1 \\
    0 & 0 & 0 & 0 & 0 & 0 & -2 & 0 \\
    2 & 0 & 0 & 1 & 0 & 0 & 0 & 0 \\
    0 & 0 & 0 & 0 & 0 & 0 & 0 & 0 \\
    0 & 0 & 0 & 0 & -1 & 0 & 0 & 0
  \end{bmatrix}, \qquad
  \vec{d} =
  \begin{bmatrix}
    2 \\ 0 \\ 0 \\ -4 \\ 0 \\ 0 \\ 0 \\ 0 
  \end{bmatrix}, \qquad
  \vec{v} =
  \begin{bmatrix}
    1 \\ -e^{-2i(\theta-\theta_{\kappa})} \\ 0 \\ 0 \\ 0 \\ 0 \\ 0 \\ 0
  \end{bmatrix}.
\end{equation}
\end{widetext}
In the second line of Eq.~(\ref{eq:Calculate_S_I(Delta)}) the
row-vector, $\vec{v}^{\mathrm{T}}$, extracts the two first entries of
$[\vec{y}(t,\tau)-\vec{y}(t,\infty)]$ and multiplies these by the
appropriate weights according to Eq.~(\ref{eq:R(t;tau)_ad_elim}). In
the third line the solution~(\ref{eq:Solution_y(tau)-y(inf)}) is used
and the $\tau$-integration of $e^{\vec{A}\tau}\cos(\Delta\tau)$ is
carried out. The last line includes (in similarity with the discussion
below Eq.~(\ref{eq:ddt_Sz_triplet_noDecay})) the re-expression of the
quadratic and cubic terms of $\vec{y}$ by the linear and quadratic
terms of $\vec{x}$ (in steady state) as a particular property of the
triplet space: $\vec{y}(t,0) = \vec{C}\vec{x}(t) + \vec{d}$.

For completeness, we present a single-qubit version of the above
results in the more general case that decoherence processes and the
Stark-shift term of Eq.~(\ref{eq:Heff}) are included. The vectors
$\vec{x}$, $\vec{y}$, and $\vec{v}$ are restricted to the first three
terms only, and the correlated-decay operator reduces to $\c_1 =
\sqrt{\gammap}\pauli_-$ presenting effectively an extra
qubit-population decay channel. The last line of
Eq.~(\ref{eq:Calculate_S_I(Delta)}) remains valid when replacing: 
\begin{equation}
  \begin{split}
  \vec{A} = 
  &\begin{bmatrix}
    -(\gammaperpeff-i\Deltaaeff) & 0 & -\frac{i\chi}{2} \\
    0 & -(\gammaperpeff+i\Deltaaeff) &  \frac{i\chi}{2} \\
    -i\chi & i\chi & -\gammapareff
  \end{bmatrix}, \\
  \vec{b} = 
  &\begin{bmatrix}
    0 \\ 0 \\ \gammapareff
  \end{bmatrix}, \quad
  \vec{C} =
    \begin{bmatrix}
      0 & 0 & \frac{1}{2} \\ 0 & 0 & 0 \\ 0 & -1 & 0
    \end{bmatrix},
  \quad \vec{d} = 
  \begin{bmatrix}
    \frac{1}{2} \\ 0 \\ 0
  \end{bmatrix},
  \end{split}
\end{equation}
where $\gammaperpeff = \frac{1}{\tau} + \frac{\gammapareff}{2}$ with
$\gammapareff = \gammapar + \gammap$, and $\Deltaaeff = \Deltaa -
\frac{\gammap\Deltaca}{2\kappa}$. The single-qubit version of
Eq.~(\ref{eq:Mean_Sx_triplet_SS}) becomes:
\begin{equation}
  \mean{\pauli_x} = -\frac{\chi\Deltaaeff}
  {(\Deltaaeff)^2 + (\gammaperpeff)^2(1+\frac{\chi^2}
    {\gammaperpeff\gammapareff})}.
\end{equation}


%

\end{document}